\documentclass[12pt]{article}
\usepackage{amssymb}
\usepackage{graphicx}
\graphicspath{ {images/} }
\usepackage{amsmath}
\usepackage{hyperref}
\usepackage{amsmath}
\usepackage{slashed}
\numberwithin{equation}{section}
\begin{document}
\title{\bf Non-Abelian two-form gauge transformation and gauge theories in six dimensions}
\author{Ahmad Moradpouri\thanks{Email: ahmadreza.moradpour@gmail.com}  \hspace{2mm}\\ 
 }
\date{\today}
\maketitle

\abstract A new non-Abelian gauge transformation for two-forms is introduced. Construction is based on a fixed map from the spacetime to the loop space which attachs a closed loop to each point of the spacetime. It is argued that this set-up is consistent with the surface ordering ambiguity which is the main problem to construct the Wilson surface operator for non-Abelian groups. With the aim of the Wilson surface operator, we achieve a non-Abelian gauge transformation for two-forms. We interpret the Dirac operator as a vector field and define a covariant derivative and rederive the gauge transformation of the two-form. At the end, we construct an Abelian interacting gauge theory in six dimensions.

%--------------------------------------------------------------------------
\section{Introduction} \label{intro}
The existence of the (2, 0) superconformal theory is one of the most incredible predictions of string theory which was based on string dualities \cite{Ref1}. From M-theory point of view, this theory describes the low energy limit of a stack of the M5-branes. The dynamics of a single five-brane in eleven- dimensional supergravity was explored in \cite{Ref2}, \cite{Ref3}. The (2, 0) superconformal theory has some interesting physical and mathematical consequences. Based on AdS/CFT correspondence, M-theory on $AdS_{7}\times S^4$ is dual to this theory and many interesting four-dimensional superconformal field theories can be achieved by compactification of this theory on a Riemann surface \cite{Ref4}. On the other hand, it is believed that this theory has some interesting relations to geometric Langlands program \cite{Ref5}, \cite{Ref6}.\\ 
There is no perturbative parameter in M-theory and correspondingly 6D-superconformal theories can not be described perturbatively. In spite of many attempts to describe this theory in Lagrangian formalism, there is no known description based on action functional. There are some reasons for this difficulty which the most important of them is the existence of a self-dual two-form in the (2, 0) tensor multiplet. Because of the self-duality, it is difficult to write a covariant action for this two-form. Furthermore, finding a generalized non-Abelian theory of the two-forms is another problem.\\

From the mathematical point of view, a non-Abelian two-form can be described by non-Abelian gerbes \cite{Ref7}, \cite{Ref8}. In this framework in addition to the two-form, we also need to introduce a new gauge field $A$. The tensor multiplet in six dimensions does not contain a vector gauge field and if this pattern plays a role in 6d-superconformal field theories, it is not clear how do we can manage the role of this gauge field which introduce extra bosonic degrees of freedom to the theory. It might be possible as like as three dimensional superconformal field theories to introduce a topological field theory for the dynamics of the gauge field $A$ to resolve this obstacle \cite{Ref9}, but as far as we know, there is no topological field theory for a gauge field in six dimensions.\\
\\ 
The aim of this paper is to find solutions for some of these problems. We consider a simpler theory that only has a Dirac spinor and a two-form and we focus on the non-Abelian generalization part of the problem. The organization of the paper is as follow: In section II, we review some basic ingredients of QFT. In section III, we will introduce a new gauge transformation to the two-form which impose a non-local structure to the theory; and we discuss the properties of the Wilson surface operator for the non-Abelian two-form. In section IV, we define a covariant derivative with respect to the gauge transformations which was mentioned in section III and we will derive a new Abelian gauge theory in six dimensions which we expect to have some similar structure to the final (2, 0) superconformal field theory in six dimensions.

%-------------------------------------------------------------------------

\section{General aspects}
The (2, 0) Tensor multiplet of superconformal algebra in six dimensions consists of five scalar $\phi_{i}(i=1,...,5)$, two Weyl spinors $\psi_{\alpha}(\alpha=1,2)$  and a self-dual two-form $B$  \cite{Ref10}. If we assume that the kinetic term has the usual squaring term for the fields, we will have

\begin{equation}\label{x1}
\begin{split}
 \mathcal{L_{\phi}}&= \frac{1}{2}\partial_{\mu}\phi\partial^{\mu}\phi,\quad\quad
 \mathcal{L_{\psi}}=i\bar{\psi}\gamma^{\mu}\partial_{\mu}\psi,
  \\
 \mathcal{L_{B}}&=-\frac{1}{6} H_{\mu\nu\rho}H^{\mu\nu\rho},\quad\quad H=dB.
 \end{split}
\end{equation}
The mass dimensions of the scalars, the spinors and the two-form in six dimensions are as follow:
\begin{equation}\label{x2}
 [\psi]=\frac{5} {2},
 [\phi]=2,
 [B_{\mu\nu}]=2.
\end{equation}
The simplest interaction of the spinors $\psi_{\alpha}$ with the B-field has the following schematic form:
\begin{equation} %\label{x3}
\propto\bar {\psi}\psi B.
\end{equation}
The mass dimensions of this interaction in six dimensions is seven and we have to introduce a parameter with negative dimensions. This parameter eliminates conformality and renormalizability of the theory. With the same analogy, the spinor field can not couple to the scalars of the theory and it seems that the spinor fields should be decoupled from the theory.\\
With the dimensional analysis, there are only few acceptable interactions:
\begin{equation} %\label{x4}
\mathcal{L}\propto B^{3},\quad\phi B^{2},\quad\phi^{3}.
\end{equation}
The above results are achieved with the assumption that the kinetic terms of the above fields have the usual form. One can imagine that there is a fixed point in $\beta$-function where the mass dimensions of these fields at fixed point have different mass dimensions. Such a phenomenon can occur in three-dimensional superconformal field theories when we look at them as the low energy limit of a stack of the $D2$-branes. In this scenario, bosonic degrees of freedom include seven scalars and one Yang-Milles vector field which live in the world-volume of the $D2$-branes. In three dimensions, Yang-Mills coupling constant has positive dimensions and the mass dimension of the Yang-Mills vector field is one half. Such a theory can not be a candidate of a conformal theory; but we can assume that there is an IR fixed point which on there, the mass dimension of the Yang-Mills vector field changes and would be equal to one. This theory can be described by a Chern-Simon theory which has not local degrees of freedom and the degree of freedom of previous vector field will shown itself as a new scalar field. This is a natural description from the M-theory point of view which is the strongly coupled limit of type $IIA$ string theory and the $D2$-branes can be mapped to the $M2$-branes; and in the low energy limit, bosonic degrees of freedom of a stack of the $M2$-branes contain eight scalar field. Is there a similar scenario for 6D-superconformal field theories?\\
Based on the reasons that explained below, it seems that this is not a reasonable assumption.\\
\\
1. Even in the Abelian case, beside the Lagrangian for the two-forms in eq. (\ref{x1}), a Lagrangian was not written for a two-form field in six dimensions. If we look at such a term in the language of forms and with the assumption that kinetic term is quadratic with respect to the two-form, the Lagrangian should be a six-form and there is only one possibility which is eq. (\ref{x1}).\\
\\
2. The most important reason might be the fact that the compactification of the (2, 0) 6D-superconformal theory on a torus should be described by $N=4$ super Yang-Mills in four dimensions in the low-energy limit \cite{Ref6}. With the unusual term for the kinetic term, it seems impossible to achieve this goal. 
If there exist a Lagrangian description of the superconformal theories in six-dimensions, there should be a non-Abelian extension formalism of the two-forms which is a necessary ingredient to construction a Lagrangian for a stack of the $M5$-branes.\\
\\
\section{The Wilson operator}
\begin{figure}
\centering
\includegraphics[width=0.4\textwidth]{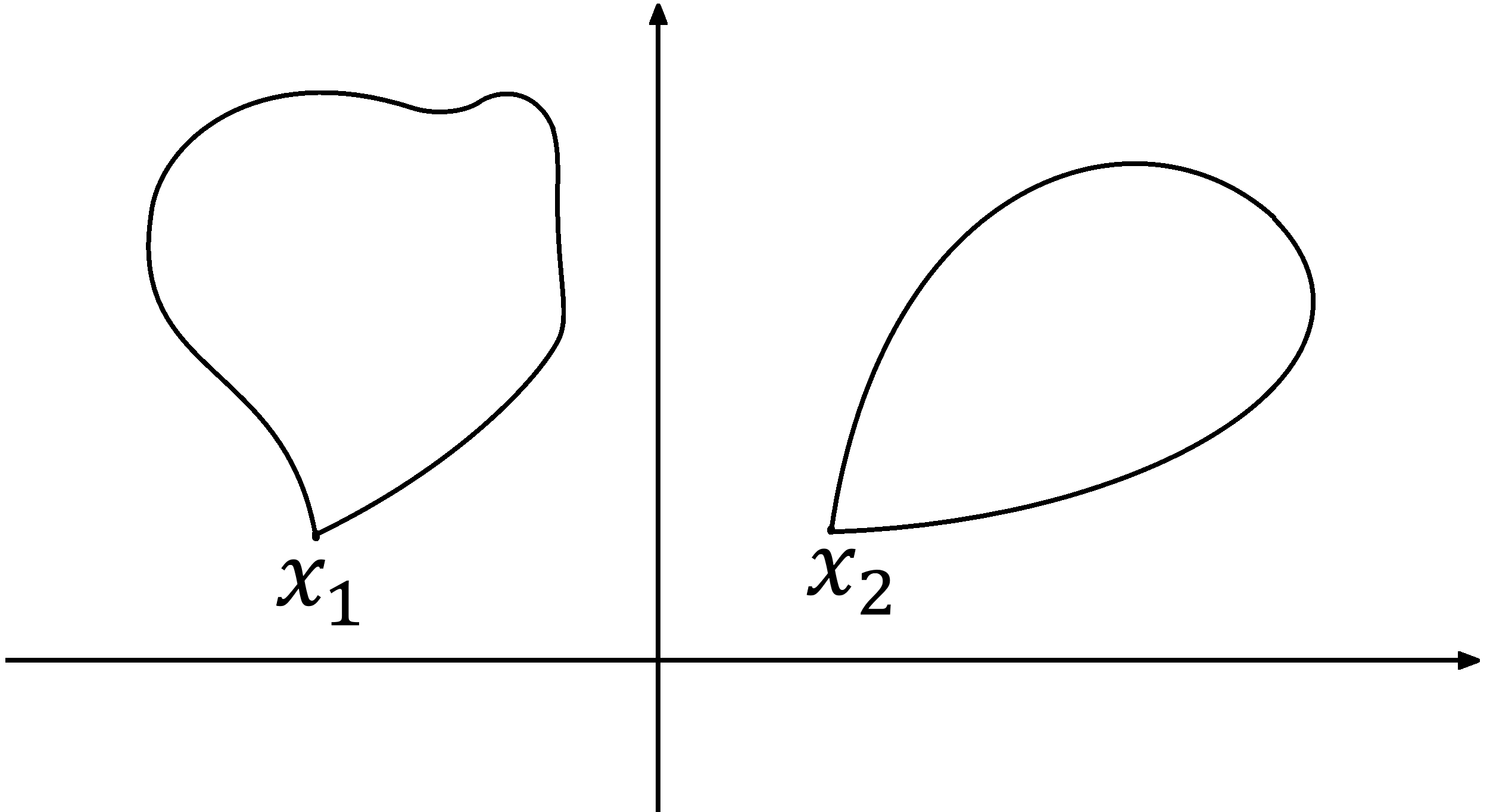}
  \caption{Closed loops which are attached to the points of the spacetime.}
  \label{fig:boat1}
\end{figure}

\begin{figure}
\centering
\includegraphics[width=0.4\textwidth]{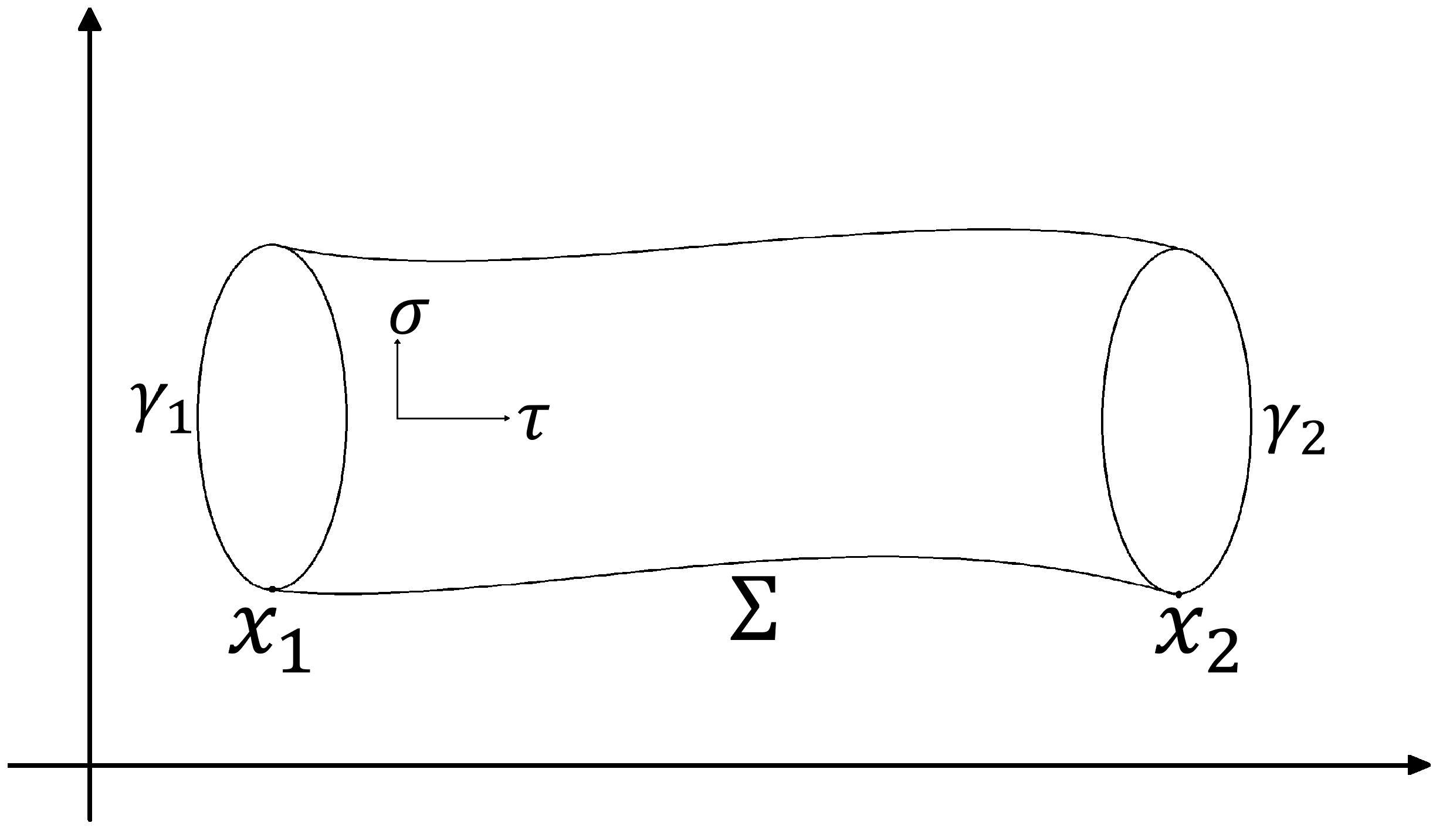}
  \caption{The surface which connects the loops $\gamma_1$ and $\gamma_2$ at the points $x_1$ and $x_2$. }
  \label{fig:boat2}
\end{figure}

\begin{figure}
\centering
\includegraphics[width=0.3\textwidth]{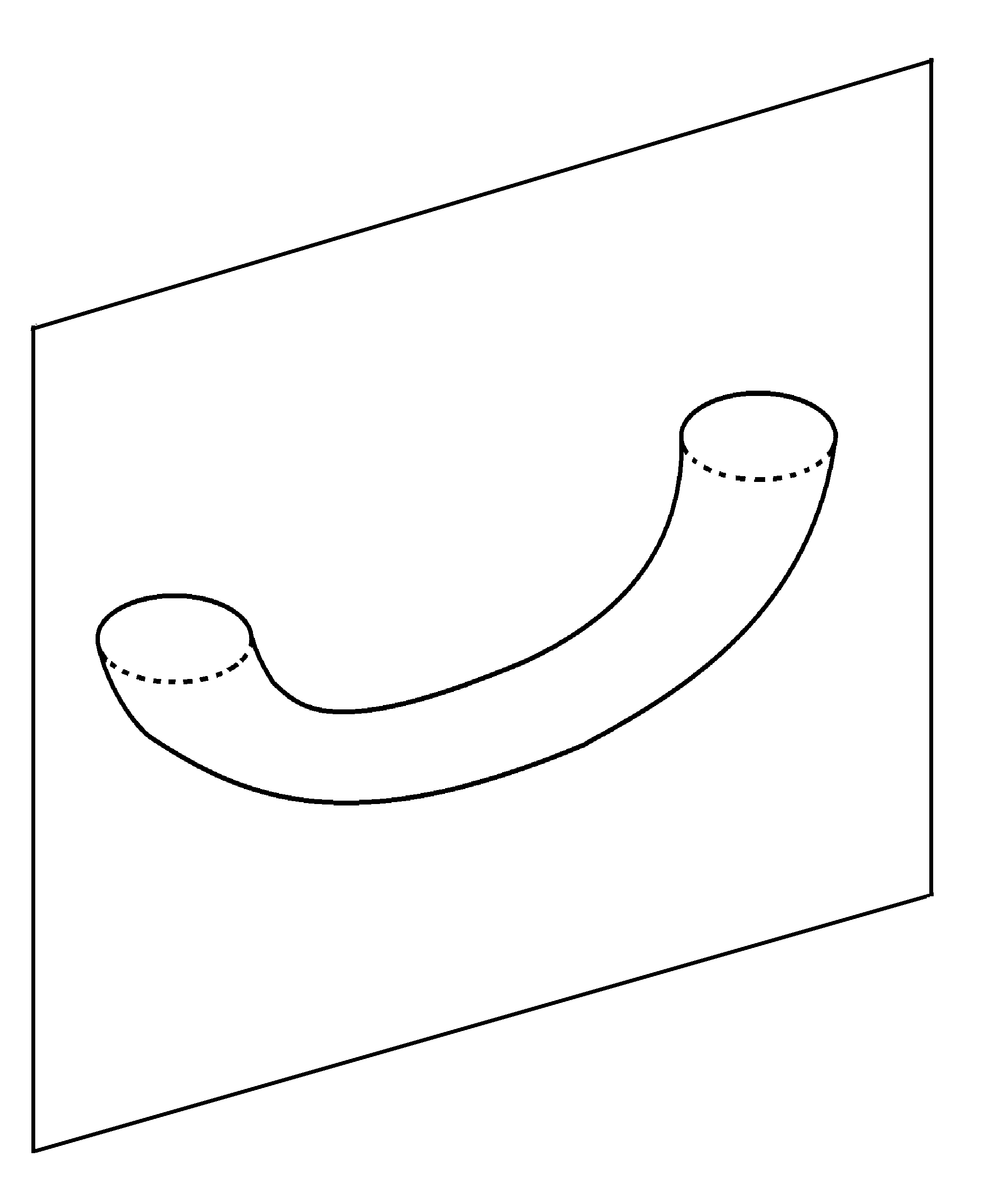}
  \caption{The intersection of an $M2$-brane and an $M5$-brane.}
  \label{fig:boat3}
\end{figure}

Yang-Mills theory, from the physical point of view, starts from a Lagrangian which have a global symmetry and by making the theory invariant under local transformations, a new gauge field will be introduced. Gauge transformation in the Abelian case has the following form:
\begin{equation} %\label{x5}
\begin{split}
\delta\psi(x)&=i\alpha(x)\psi(x),\\
\delta A_{\mu}&=\partial_{\mu}\alpha.
 \end{split}
\end{equation}
The second relation can be written as $\delta A=\delta A_{\mu}dx^{\mu}=d\alpha$ and global symmetry is given by the solutions of the eq. $d\alpha =0$ which are constant functions on manifold. In analogy of these results, the gauge transformation of a two-form that will be obtained is $\delta B=dU$ which $U$ is a one-form and global symmetry is given by the solutions of the eq. $dU=0$. In a manifold with $b_{1}\neq0$, in addition to trivial solution which $U$ is exact, there are some non-trivial solutions to this condition. In our toy model which contains a two-form and a spinor, the exactness of $U$ leads to the trivial gauge transformation for the two-form and the spinor($\delta \psi=0,\delta B=0$).\\
One of the most important operators in the gauge theories is the Wilson line operator
 \begin{equation} %\label{x6}
 W=exp(i\int_{x_{1}}^{x_2}A_{\mu}dx^{\mu}),
\end{equation}
which has a nice gauge transformation (We have written the Abelian transformation below).
\begin{equation}\label{x7}
\begin{split}
W'&=exp(i\int_{x_{1}}^{x_2}A'_{\mu}dx^{\mu})=W exp(i\alpha(x_2)-i\alpha(x_{1}),\\
&=exp(i\alpha (x_2)) W exp(-i\alpha(x_1)),\quad W'=PWP^{\dagger},
\end{split}
\end{equation}
where $P$ is defined by $P=exp(i\alpha(x))$.
As an equivalent way, we can achieve the gauge transformation of the gauge field $A$ by postulating the transformation of the Wilson line operator in eq. (\ref{x7}).\\  
In the Abelian case, to the toy model that we have considered, the most natural extension of the transformations in eq. (\ref{x5}) are as follow:\\
\begin{equation}\label{x8}
\begin {split}
\delta\psi(x)&=\left( i\oint_{x} U_{\mu}(t)dt^{\mu} \right) \psi(x),\\
\delta B&=dU,
\end{split}
\end{equation}
where $U$ is a one-form field which is integrated over a loop around the reference point $x$ (Figure \ref{fig:boat1}).\\
In eq. (\ref{x8}), we postulate the first relation and trying to obtains the second relation as a consequence of the first relation.\\
Rather than the transformation of the spinor field in eq. (\ref{x8}), one can also consider another transformation such as
\begin{equation}\label{x9}
\delta\psi(x)=i\oint_{x} U_{\mu}(t)\psi(x-t)dt^{\mu}.
\end{equation}
Here,($t^\mu=t^\mu(\lambda),t^\mu(0)=x^{\mu} $). In eq. (\ref{x9}), the transformation of the spinor field $\psi$ at the point $x$ depends on the value of $\psi$ along the curve which is unnatural and working with this transformation does not lead to acceptable results.
The first relation of eq. (\ref{x8}) is logical from the Wilson-surface-operator point of view in the Abelian case. The Wilson surface operator is defined by
\begin{equation}\label{x10}
W=exp(i\int_{\gamma_1}^{\gamma_2}B),
\end{equation}
and it relates the transformation of $\psi$ over loop $\gamma_1$ at the point $x_1$ to the transformation of $\psi$ over loop $\gamma_2$ at the point $x_2$. The closed loops $\gamma_1$ and $\gamma_2$  are the boundaries of the surface $\Sigma$ where the two-form $B$ is integrated over it( Figure \ref{fig:boat2}).\\
The gauge transformation of the Wilson surface operator can be simplified by as follow: 
\begin{equation}\label{x11}
W'=exp(i\int B')=exp(i \int(B+dU))=W exp(i\int_{\Sigma} dU),\\
\end{equation}
and by using the Stokes theorem,
\begin{equation}\label{x12}
\int_{\Sigma} dU=\int_{\partial \Sigma}U=\int_{\gamma_{2}}U-\int_{\gamma{1}}U,\quad\ W'=exp(i\int_{\gamma_2}U)Wexp(i\int_{-\gamma_1}U),
\end{equation}
which confirms  the transformation of the spinor $\psi$ in eq. (\ref{x8}). By considering the transformation of the $\psi$ and the Wilson surface operator, it is trivial that we can obtain the transformation of the B-field.\\
The loops have interesting interpretation. They are the closed strings which will be obtained from the intersection of the cylindrical $M2$-branes and the $M5$-branes. ( Figure \ref{fig:boat3})\\
\\
\subsection{Surface ordering problem}
In the non-Abelian case, because of the difficulty to define surface ordering, the Wilson surface operator is not well-defined. The basic idea which is explored in \cite{Ref11} and \cite{Ref12} is as follow: We are using the notations of \cite{Ref11} and \cite{Ref12} to explain the idea. A two-form should be coupled to a string and it sweeps a two-dimensional surface in the spacetime $x^{\mu}(\tau,\sigma)$. The wave functional of this system which is defined over a configuration space where each point of it is a closed curved is denoted by $\varPhi(x(\sigma))$. The non-Abelian gauge transformations act on $\varPhi^{A}(x(\sigma))$ by\\
\begin{equation}\label{x13}
\varPhi^{A}(x(\sigma))\longmapsto T^{A}_{B}(x(\sigma))\varPhi^{B}(x(\sigma)).
\end{equation}
Here, $\varPhi^{A}(x(\sigma))$ is the representation of the gauge group G and $T^{A}_{B}(x(\sigma))$ are matrices which are associated to the representation. To achieve gauge invariance, one needs to define a connection over the configuration space $\varPhi(x(\sigma))$ which allows us to transport the field $Phi(x(\sigma))$ along a two-dimensional surface $x^{\mu}(\tau,\sigma)$. The parallel transport equation is given by
\begin{equation}\label{x14}
\frac{d\varPhi(\tau)}{d \tau}=-A(\tau)\varPhi(\tau).
\end{equation}
Here, $A=A^{a}t_{a}$ is a G-valued one-form in $x(\sigma)$-space and $A^{a}$ is defined by\\
\begin{equation}\label{x15}
A^{a}=-\int d\sigma A^{a}_{\mu\nu}\frac{\partial x^{\mu}}{\partial\sigma}\frac{\partial x^{\nu}}{\partial\tau},
\end{equation}
The key point is that a two-dimensional surface which is parameterized by another coordinates should describe the same physics. The path independence of the theory imposes some restrictions to the algebra of the surface deformation generators at the equal time \cite{Ref13}
\begin{equation}\label{x16}
\begin {split}
[\mathcal{R_{\perp}(\sigma)},\mathcal{R_{\perp}(\sigma')}]&=(\mathcal{R}^{1}(\sigma)+\mathcal{R}^{1}(\sigma'))\delta_{,1}(\sigma-\sigma')\\
[\mathcal{R}_{1}(\sigma),\mathcal{R_{\perp}(\sigma')}]&=\mathcal{R}_{\bot}(\sigma)\delta_{,1}(\sigma-\sigma')\\
[\mathcal{R}_{1}(\sigma),\mathcal{R}_{1}(\sigma')]&=(\mathcal{R}_{1}(\sigma)+\mathcal{R}_{1}(\sigma'))\delta_{,1}(\sigma-\sigma').
\end{split}
\end{equation}
Here, $\mathcal{R_{\perp}}$ and $\mathcal{R}_{1}$ are the generators of the surface deformation and $\mathcal {R}^1=g^{11}\mathcal{R}_{1}$ where $g^{11}$ being the metric on the $\tau=cte$ strings. Decomposition of $\frac{\partial x^{\mu}}{\partial\tau}$ in terms of normal and tangential components is given by  
\begin{equation}\label{x17}
\frac{\partial x^{\mu}}{\partial \tau}=\xi^{\bot}n^{\mu}+\xi^{1}\frac{\partial x^{\mu}}{\partial\sigma}.
\end{equation}
Here, $n^{\mu}$ is the unit normal to the $\tau=cte$ closed curves. $A$ In eq. (\ref{x14}) can be written as: 
\begin{equation}\label{x18}
A=\int d\sigma (\xi^{\bot}\mathcal{R_{\bot}}+\xi^{1}\mathcal{R}_1),
\end{equation}
where $\mathcal {R_{\bot}}=-A^{a}_{\mu\nu}\frac{\partial x^{\mu}}{\partial\sigma}n^{\nu}$ and $\mathcal{R}_{1}=0$. The universal algebra in eq. (\ref{x16}) takes the simple form
\begin{equation}\label{x19}
[\mathcal{R_{\perp}(\sigma)},\mathcal{R_{\perp}(\sigma')}]=0,
\end{equation}
thus $A_{\mu\nu}$ must be Abelian.\\

\subsection{Fixed map and the non-Abelian gauge transformations}
In spite of these difficulties, the first transformations in eq. (\ref{x8}) have a different interpretation. We have attached specific closed curves at each point of the spacetime which are fixed closed curves and these loops should not be changed under changing of parameterization. Different attaching of the curves can be shown by different maps from the spacetime to loop space\\
\begin{equation}\label{x20}
\begin{split}
\varOmega:spacetime&\longmapsto loopspace\\
          x&\longmapsto\gamma_{x}
\end{split}        
\end{equation}

\begin{figure}
   \centering
  \begin{minipage}{0.4\textwidth}
    \includegraphics[width=\textwidth]{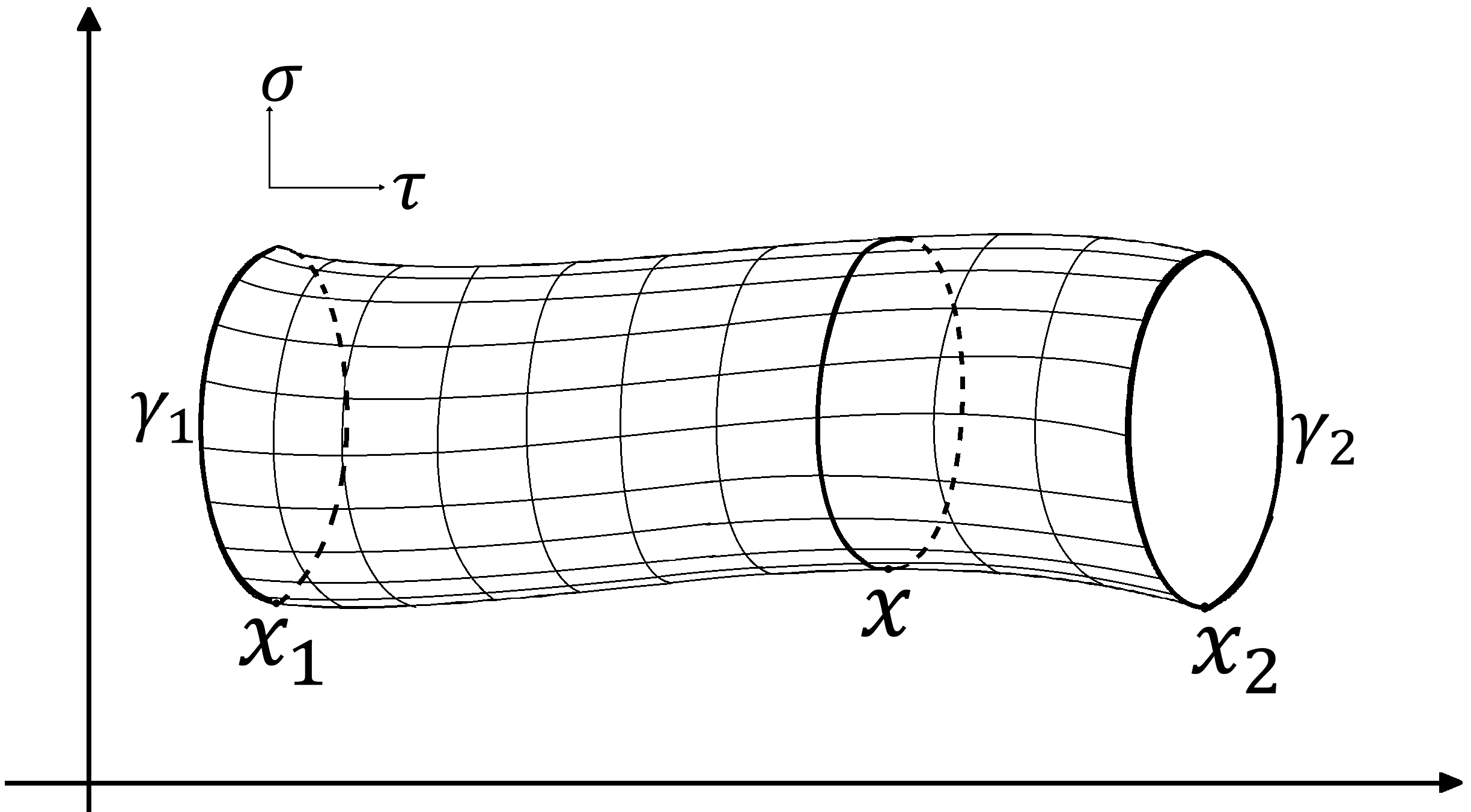}
    \caption{The closed loops in the coordinates $(\sigma, \tau)$}
    \label{fig:boat4}
  \end{minipage}
     \centering
  \begin{minipage}{0.4\textwidth}
    \includegraphics[width=\textwidth]{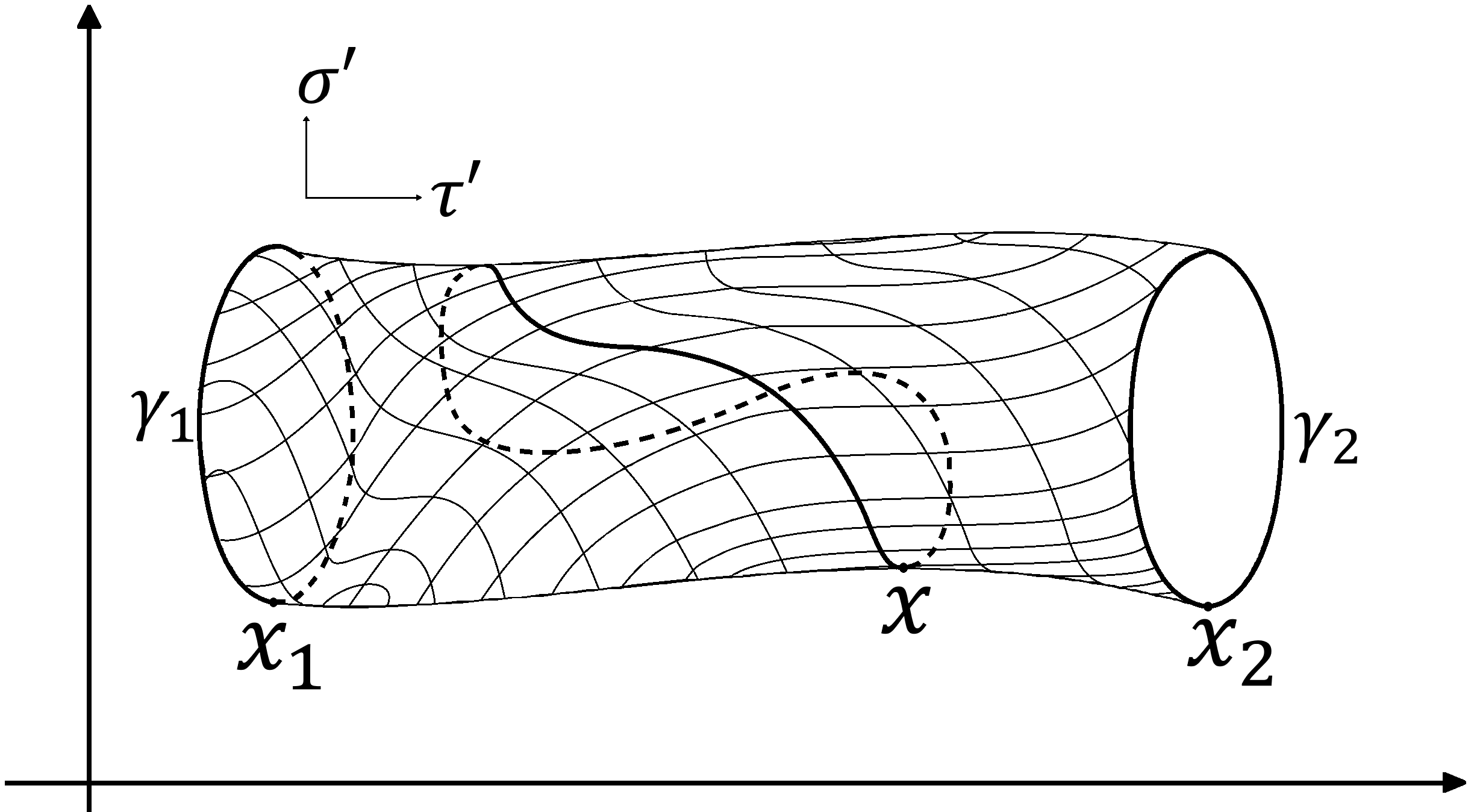}
    \caption{The closed loops in the coordinates $(\sigma', \tau')$}
    \label{fig:boat5}
  \end{minipage}
\end{figure}

Here, $\gamma_{x}$ is a loop attached to the spacetime at the point $x$. Consider a two-dimensional surface which connects the loop $\gamma_{1}$ at the point $x_{1}$ to the loop $\gamma_{2}$ at the point $x_{1}$ (Figure \ref{fig:boat4}). The surface that relates these two loops is not so flexible and because we have attached specific loops at each point by the map $\Omega$, it is sufficient to specify the curve $x^{\mu}(\tau,0)$  between $x_{1}$ and $x_{2}$ to determine the whole surface. If we try to change the coordinates of the surface, the loops are deformed to different loops in the new coordinates at the same point in the spacetime; for example, the loops $\gamma_{x}$ and $\gamma'_{x}$  at the point $x$ which have shown by bold lines in figures \ref{fig:boat4} and \ref{fig:boat5}.\\
\\
As a result, it is not compatible with our fixed map $\Omega$, therefore the map $\Omega$ defines a unique coordinates on the above surface. Our strategy is as follow: first we define the gauge transformations for a fixed map $\Omega$ and we try to find a Lagrangian that is invariant under the transformations and at the end, we define an effective action by integrating over all maps from the spacetime to the loop space.\\ 
Working with a fixed map, one can show that if the Wilson surface operator transforms according to $W'=PWP^{\dagger}$ where $P$ is given by $P=\mathcal{P_{\sigma}}exp(i\oint_{x}U)$ which $\mathcal{P_{\sigma}}$ is path ordering operation along the constant $\tau$ closed curves, the gauge transformation of the B-field can be achieved as follow:
\begin{equation}\label{x21}
\delta B=dU+i[\oint_{x}U,B],
\end{equation}
where the non-Abelian B-field is in the adjoint representation of the Lie group G and $U$ is given by $U=U_{\mu}^{a}T^{a}dx^{\mu}$.
The relation (\ref{x21}) also will be achieved from a covariant derivative that will be defined in the following. 
One can ask: Is the Dirac Lagrangian invariant under the symmetry
$\delta\psi(x)=i\oint_{x} U_{\mu}(t)dt^{\mu}\psi(x))$?
The variation of the Dirac Lagrangian is as follow:\\
\begin{equation}\label{x22}
\delta\mathcal{L}=\delta\bar \psi(x)i\gamma^{\mu}\partial_{\mu}\psi (x)+\bar \psi (x)i\gamma^{\mu}\partial_{\mu}\delta\psi (x)=\bar \psi (x)\gamma^{\mu}\psi (x)(i\partial_{\mu}\oint_{x}U).
\end{equation}
Partial derivatives are not meaningful for calculating the change of $\oint_{x}U$ with respect to the reference point $x$. $\oint_{x}U$ in addition to the point $x$ depends on other points of the curve and is not a local function and we need some information along the curve. With the infinitesimal change of the reference point $x$, it is not clear that the other points also should change equally. We can define a vector field on the spacetime manifold and try to replace partial derivatives with the Lie derivatives; but is there a reference vector field on the manifold?\\
The Dirac operator $\slashed\partial=\gamma^{\mu}\partial_{\mu}$ is the natural guess as a vector field. Although $\gamma^{\mu}$ are matrices, there are some similarities with a usual vector field. We can define integral curves of the Dirac operator as an eigen-value problem $\frac{dx^{\mu}}{dt}\chi=\gamma^{\mu}\chi$. The Lie derivative of a spinor field is given by
\begin{equation}\label{x23}
\mathcal{L}_{V}\psi=V^{\mu}\nabla_{\mu}\psi-\frac{1}{8}\nabla_{[{\mu}}V_{\nu]}\gamma^{[\mu}\gamma^{\nu]}\psi.
\end{equation}\\
We define the vector field $V$ by $V=\gamma^{\mu}\partial_{\mu}=e^{\mu}_{a}\gamma^{a}\partial_{\mu}=V_{a}\gamma^{a}$ and it is reasonable to assume that the Lie derivative of the spinor $\psi$ along this vector field can be defined by
\begin{equation}\label{x24}
\mathcal{L}_{V}\psi=\mathcal{L}_{V_{a}\gamma^{a}}\psi=\gamma^{a}\mathcal{L}_{V_{a}}\psi.
\end{equation}
At the moment, we ignore the local Lorantz transformations; so the Lie derivative of spinor $\psi$ can be simplified to $\mathcal{L}_{\slashed\partial}\psi=\gamma^{\mu}\partial_{\mu}\psi$; so the Dirac Lagrangian can be written as $\mathcal{L}=i\bar\psi\mathcal{L}_{\slashed\partial}\psi$. As a result, the variation of the Dirac Lagrangian with respect to the first transformation of the eq. (\ref{x8}) can be simplified as follow:
 \begin{equation}\label{x25}
 \delta\mathcal{L}=i\bar\psi(x)(i\mathcal{L}_{\slashed\partial}\oint_{x} U)\psi(x)=-i\bar\psi(x)\left( i\oint_{x}\gamma ^{\nu}(\partial_{\mu} U_{\nu}-\partial_\nu U_{\mu}) dt^\mu \right) \psi(x).
 \end{equation}
We can define a covariant derivative that act on the spinor field in such a way to cancel this term, which have the following form:
 \begin{equation}\label{x26}
 D_{\slashed\partial}\psi(x)=\mathcal{L}_{\slashed\partial}\psi(x)+i(\oint_{x}B_{\mu\nu}\gamma^{\nu}dx^{\mu})\psi(x).
 \end{equation}
If we impose that the covariant derivative transform according to $\delta D_{\slashed\partial}\psi(x)=(i\oint_{x}U)D_{\slashed\partial}\psi (x)$, the gauge transformation of the two-form gauge field will be find as follow:
 \begin{equation}\label{x27}
 \delta B_{\mu\nu}=\partial_{\mu}U_{\nu}-\partial_{\nu}U_{\mu}+i[\oint_{x}U, B_{\mu\nu}].
\end{equation}
This is the gauge transformation of the non-Abelian gauge field which is the generalization of the Abelian transformation and was obtained by the Wilson surface operator previously in (\ref{x21}).
The construction of a field strength which has the nice gauge transformation due to gauge transformation (\ref{x27}) is the next step to construct an interacting gauge theory. Unfortunately, at the moment, we have not found a suitable non-Abelian generalization of the Abelian field strength; but with the help of the Abelian field strength, we can construct a QED-like gauge theory in six dimensions which have not a dimensionful parameter. The Lagrangian of this theory is given by
\begin{equation}\label{x28}
S=\int dx^{6}[i\bar\psi D_{\slashed\partial}\psi-\frac{1}{6}H_{\mu\nu\rho}H^{\mu\nu\rho}].
\end{equation}
The above action is defined only for a fixed map from the spacetime to the loop space.  We can achieve an effective action by integrating over all maps. This is closely related to integrating over fermions in QED path-integral which defines an effective action for the Maxwell gauge field.\\
\\
In this case, the effective action can be defined as follow: 
\begin{equation}\label{x29}
exp(-S_{eff})=\int {D} {\Omega} {exp(-S)}.
\end{equation}
the above effective action is invariant under the generic map from the spacetime to the loopspace. It is notable to note that the action (\ref{x29}) is an Abelian action and trying to find a non-Abelian extension of the field strength of the B-field is a necessary ingredient to the construction of the non-Abelian interacting gauge theory. Furthermore, we can introduce a coupling constant to the Wilson surface operator (\ref{x10}) and the covariant derivative (\ref{x26}). (2, 0) Superconformal theory has not adjustable parameters and if this pattern works, this coupling constant shouldn’t be present in the theory but we have only focused on a toy model which is not supersymmetric and it might be possible that the coupling constant would be fixed in supersymmetric theories.  These problems can be addressed in the following works.\\
\\
\section{Conclusion}
We have introduced a toy model which contains a two-form and a Dirac spinor.  The ambiguity to define the Wilson surface operator is explored and we defined a fixed map from the spacetime to the loop space to solve this obstacle. This map determined a specific coordinates to the surface which connects the loops which are specified from the fixed map  in the different points of the spacetime. With the aim of this map, we have derived a non-Abelian gauge transformation of two-forms. We also defined a covariant derivative with respect to the two-forms and constructed an interacting Abelian gauge theory in six-dimensions. By integrating over all maps we have obtained an effective field theory.\\  

%--------------------------------------------------------------------------

\end{document}